# Highly efficient MRI through multi-shot echo planar imaging

Congyu Liao[a,b], Xiaozhi Cao[c], Jaejin Cho[a,b], Zijing Zhang[a,d], Kawin Setsompop[a,b], Berkin Bilgic[*a,b]

[a]Athinoula A. Martinos Center for Biomedical Imaging, Massachusetts General Hospital, Charlestown, MA, USA; [b]Department of Radiology, Harvard Medical School, Boston, MA, USA; [c]Center for Brain Imaging Science and Technology, Zhejiang University, Hangzhou, China; [d]The State Key Laboratory of Modern Optical Instrumentation, College of Optical Science and Engineering, Zhejiang University, Hangzhou, China

## ABSTRACT

Multi-shot echo planar imaging (msEPI) is a promising approach to achieve high in-plane resolution with high sampling efficiency and low $T_2^*$ blurring. However, due to the geometric distortion, shot-to-shot phase variations and potential subject motion, msEPI continues to be a challenge in MRI. In this work, we introduce acquisition and reconstruction strategies for robust, high-quality msEPI without phase navigators. We propose Blip Up-Down Acquisition (BUDA) using interleaved blip-up and -down phase encoding, and incorporate $B_0$ forward-modeling into Hankel structured low-rank model to enable distortion- and navigator-free msEPI. We improve the acquisition efficiency and reconstruction quality by incorporating simultaneous multi-slice acquisition and virtual-coil reconstruction into the BUDA technique. We further combine BUDA with the novel RF-encoded gSlider acquisition, dubbed "BUDA-gSlider", to achieve rapid high isotropic-resolution MRI. Deploying BUDA-gSlider with model-based reconstruction allows for distortion-free whole-brain 1mm isotropic $T_2$ mapping in ~1 minute. It also provides whole-brain 1mm isotropic diffusion imaging with high geometric fidelity and SNR efficiency. We finally incorporate sinusoidal "wave" gradients during the EPI readout to better use coil sensitivity encoding with controlled aliasing.

## 1. INTRODUCTION

Echo planar imaging (EPI) is an efficient acquisition technique for fast MRI. The single-shot variant of EPI (ssEPI) which acquires the entire k-space data in a single TR is the most commonly used approach, for its fast acquisition and its immunity to bulk motion. However, the associated lengthy echo train of ssEPI induces $T_2^*$ blurring and geometric distortion, imposing a limitation on the achievable resolution. Multi-shot EPI (msEPI) allows high-resolution imaging with reduced distortion, but combining shots is prohibitively difficult because of shot-to-shot phase variations. These variations can be mitigated using navigators[1,2], albeit at the cost of imaging efficiency and potential remaining artifacts. Navigator-free approaches[3–5] employ parallel imaging (PI)[6,7] to reconstruct each shot, from which phase variations are estimated. This imposes a limit on the distortion reduction since PI breaks down beyond $R_{inplane}>4$ acceleration.

Recently, Hankel structured low-rank constrained PI approaches[8–12] were proposed to enable navigator-free msEPI acquisition with higher in-plane acceleration to minimize geometric distortion. However, achieving adequate image quality at higher acceleration necessitates a larger number of shots (e.g., $R_{inplane}=8$ with 4-shots), thereby limiting the efficiency of such acquisitions. In previous work[13–15], we incorporated virtual coil (VC) concept[16] and simultaneous multi-slice (SMS) acquisition[17,18] into structured low-rank model and deep learning-based reconstructions to further accelerate msEPI. Similar works were reported by other groups[19,20].

In this study, we propose an efficient Blip Up-Down Acquisition (BUDA), where a 2-shot EPI sampling was performed with interleaved blip-up and -down acquisitions, and then combined with $B_0$ forward-modeling and structured low-rank reconstruction to yield distortion-free images. With BUDA, the required number of shots was reduced to two, which improved the sampling efficiency of msEPI. We demonstrate its extension to the VC concept through structured low-rank reconstruction and obtain high-quality images from a partial Fourier 75% acquisition. We further combine BUDA with a

---

novel volumetric RF-encoding technique, dubbed generalized slice dithered enhanced resolution (gSlider) acquisition [21], to enable high signal-to-noise ratio (SNR) efficiency, high isotropic-resolution imaging with high geometric fidelity. Finally, we apply our proposed BUDA-gSlider to encoding-intensive diffusion-weighted imaging (DWI) and quantitative $T_2$ mapping, to achieve rapid whole-brain 1mm isotropic DWI and $T_2$ maps without geometric distortion. In addition, we demonstrate that employing sinusoidal "wave" gradients during the EPI readout improves image quality by better harnessing coil sensitivity encoding.

## 2. METHODS

### 2.1 Multi-shot EPI with Blip Up-Down Acquisition (BUDA)

Figure 1(a) shows the sampling strategy of a two-shot BUDA acquisition, where two interleaved shots with blip-up and -down phase encoding polarity were collected. Figure 1(b) shows the flowchart of BUDA reconstruction, which includes the following steps: (i) individual SENSE reconstructions for blip-up and blip-down acquisitions, respectively. As blue arrows pointed in Figure 1(b), each individual shot has significant geometric distortion along the phase-encoding direction. (ii) Estimating the field maps using SENSE reconstructed blip-up and blip-down images. The field maps are estimated using FSL TOPUP [22,23]. (iii) Incorporating field map forward-modeling into Hankel structured low-rank constrained joint reconstruction, which can be expressed as:

$$min_x \sum_{t=1}^{N_s} \|F_t E_t C x_t - d_t\|_2^2 + \lambda \|\mathcal{H}(x)\|_*, \qquad (1)$$

where $F_t$ is the undersampled Fourier operator in $t^{th}$ shot, $E_t$ is the estimated off-resonance information, $C$ are the ESPIRiT coil sensitivities [24] estimated from a distortion-free gradient-echo prescan data, $x_t$ is the distortion-free image and $d_t$ are the k-space data for shot t. In the forward model, the distortion-free images are multiplied by coil sensitivities and distorted by the phase modulation of the off-resonance map in hybrid (x-$k_y$) space. The constraint $\|\mathcal{H}(x)\|_*$ enforces low-rank prior on the block-Hankel representation of the multi-shot data $x$, which is applied in k-space where 7x7 blocks of k-space are concatenated from each shot in the column axis.

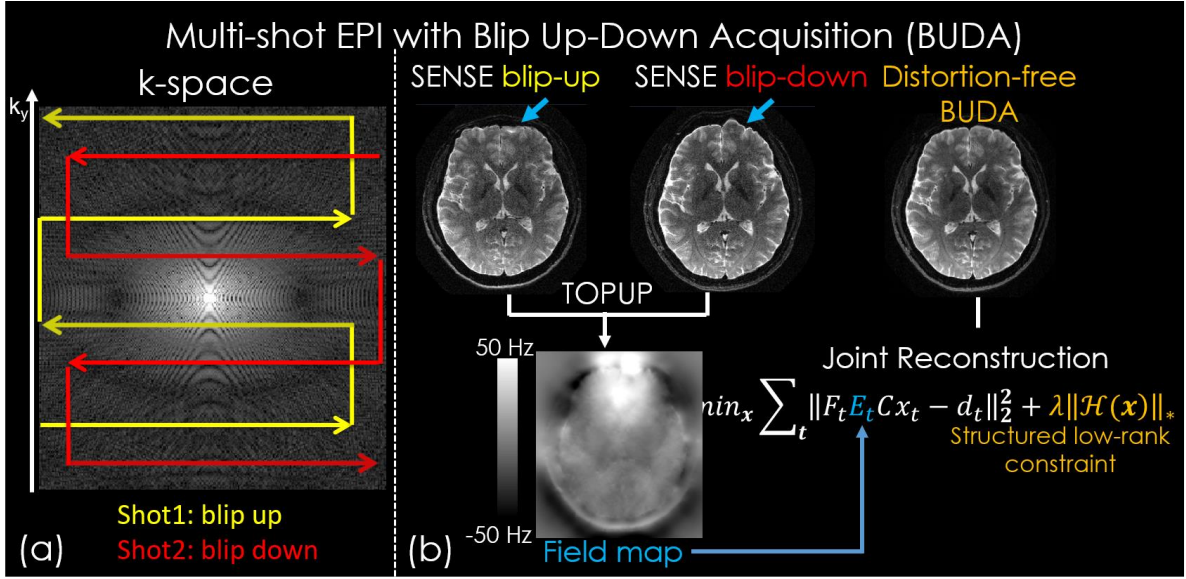

Figure 1. (a) two-shot EPI with blip up-down acquisition (BUDA) strategy. (b) The flowchart of BUDA reconstruction.

### 2.2 Joint virtual coil reconstruction for simultaneous multi-slice BUDA

The VC concept [16,25] is an approach that utilizes smooth phase prior for reconstruction. The virtual coils containing conjugate symmetric k-space signals are generated from actual coils, which effectively doubles the number available channels. These allow for incorporating additional phase information into standard SENSE[6] or GRAPPA[7] to improve the reconstruction performance in a number of applications[15,25–28]. In particular, our previous studies[10,14,15] have demonstrated that the VC reconstruction can provide improved SNR with reduced g-factor in diffusion-weighted spin-echo and spin-and-gradient echo (SAGE) EPI acquisitions. With VC reconstruction, Eq. (1) can be modified as:

$$min_m \sum_{t=1}^{N_s} \left\| \begin{bmatrix} F_t E_t C \\ F_{-t} E_t C^* \end{bmatrix} x_t - \begin{bmatrix} d_t \\ d_{-t}^* \end{bmatrix} \right\|_2^2 + \lambda \|\mathcal{H}(x)\|_* \quad (2)$$

Here, $d_{-t}^*$ is the VC k-space, and $C^*$ are the corresponding conjugate sensitivities. Figure 2 shows the flowchart of the proposed joint VC reconstruction, where the conjugate shots are incorporated into the Hankel low-rank matrix, whereby the sampled conjugate-symmetric k-space helps estimate the missing data and substantially improves the reconstruction. To better take advantage of the conjugate property of k-space, a shifted-$k_y$ EPI sampling strategy[14] ($\Delta k_y$ shift =1 for the blip-up shot and $\Delta k_y$ shift =3 for the blip-down shot at $R_{inplane}$ =4) was used to provide g-factor improvement over a non-shifted symmetric undersampling.

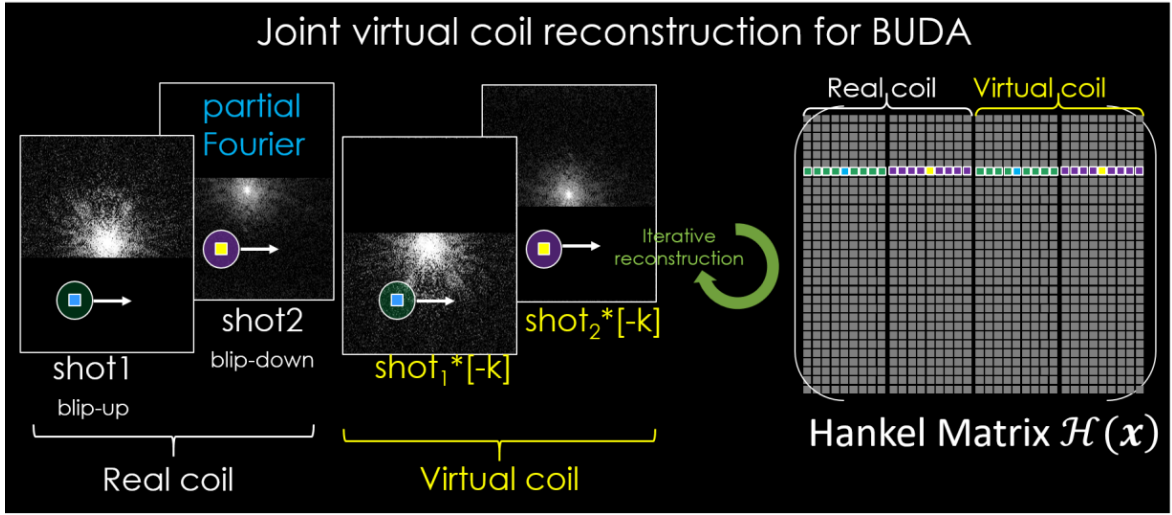

Figure 2. The flowchart of joint virtual coil reconstruction for BUDA. Virtual coils provide complementary k-space information in the presence of partial Fourier undersampling.

We further extended our VC-BUDA to simultaneous multi-slice (SMS) acquisition. Our previous work[29] developed an approach to allow structured low-rank constrained reconstruction to work with SMS encoding using the readout-extended FOV concept. This represents SMS as undersampling in the $k_x$ axis by concatenating the two slices along the readout direction. In-plane and slice acceleration could thus be captured using the Fourier operator $F_t$ in Eq. (1), now with simultaneous $k_x$-$k_y$ undersampling, which allowed us to further push the total acceleration and reduce TR for msEPI.

All in vivo measurements were performed on a 3T scanner (MAGNETOM Prisma, Siemens Healthineers, Erlangen, Germany). To validate the proposed VC-BUDA reconstruction, whole-brain 1.0×1.0×3 mm³ resolution SE-EPI data were acquired using a 32-channel head array with TE/TR=56/2700ms. 2-shots were collected at multi-band factor=2, $R_{inplane}$=4 and partial Fourier (PF) 75%. The field-of-view (FOV) of EPI acquisition is 220 mm. A FOV-matched gradient-echo (GRE) sequence was acquired to obtain distortion-free sensitivity maps for BUDA reconstruction. A fast spin-echo (FSE) data was acquired to serve as a distortion-free reference. To demonstrate the utility of our proposed method, BUDA was compared with hybrid-SENSE[51] on the same dataset.

## 2.3 Combining BUDA with RF-encoding acquisition (BUDA-gSlider) for diffusion imaging

BUDA provides a robust acquisition and reconstruction framework to achieve distortion-free high in-plane resolution MRI. To achieve high isotropic-resolution MRI with thin slices, improving the SNR efficiency of EPI acquisition is critical, especially in SNR-starved applications such as diffusion-weighted imaging (DWI). Three dimensional multi-slab diffusion-weighted EPI has emerged as a promising strategy to enhance the SNR in such acquisitions[30–32]. However, slab-boundary artifacts are key challenges for efficient sampling of whole-brain high isotropic-resolution DWI with this technique[33,34]. Another promising approach for high-SNR efficiency, high-resolution DWI is the Generalized SLIce Dithered Enhanced Resolution (gSlider) method[21]. gSlider is a simultaneous multi-slab acquisition technique with self-navigated RF slab-encoding, which has been demonstrated for motion-robust, high-resolution DWI[35]. Here we combine BUDA and gSlider to achieve high geometric fidelity, high isotropic-resolution MRI without phase navigators.

Figure 3(a) shows the sequence diagram of the proposed BUDA-gSlider, where both blip-up and blip-down shots are acquired in each RF-encoding sequentially. Five RF-encoding pulses with the sharp sub-slice encoding performance shown in Figure 3(b) were designed by Shinnar-Le Roux (SLR) algorithm[36] and used for slab-encoding[37]. With five gSlider RF-encoding pulses and 2 blip up-down shots, a total of 5×2 =10 shots were acquired for each slab. These RF excitations were then combined with blipped-CAIPI technique (multiband factor of 2)[18] to acquire 10 simultaneous slices per EPI-shot.

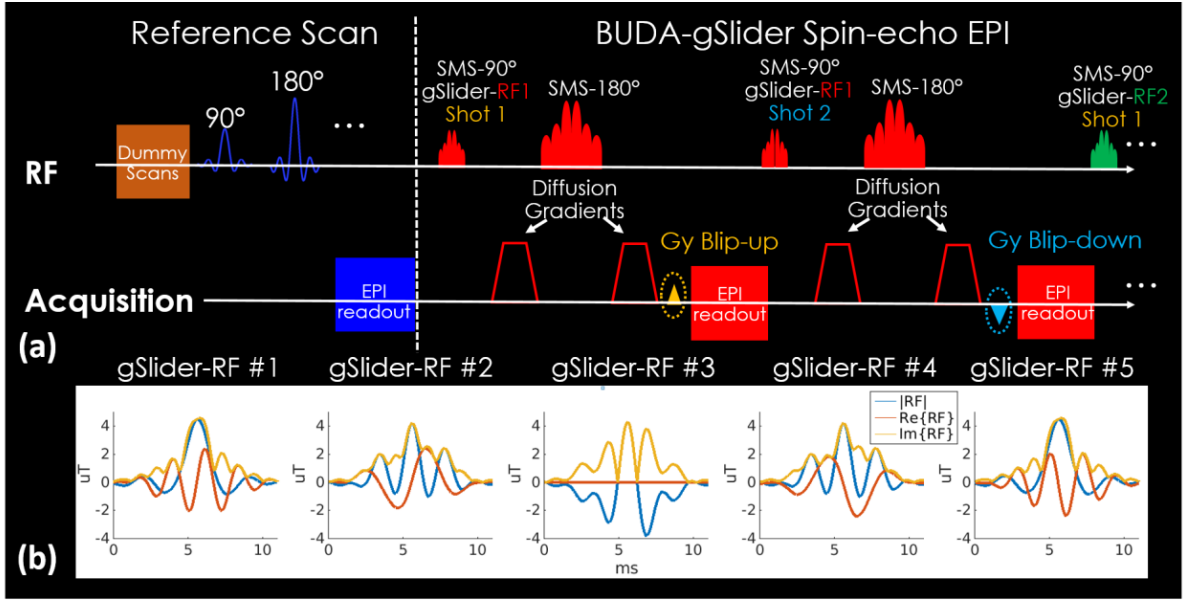

Figure 3. (a) The sequence diagram of BUDA-gSlider, where both blip-up and blip-down shots are acquired in each RF-encoding. (b) Five RF-encoded gSlider pulses used in the BUDA-gSlider sequence. 5x-gSlider with slice phase–dither encoding provides highly independent basis, while maintaining high image SNR in each individual slab acquisition.

The reconstruction of BUDA-gSlider data includes the following steps: (i) SMS-BUDA reconstruction was performed for each RF encoding to obtain distortion-free thin-slab volume. (ii) To eliminate shot-to-shot background phase variations in the acquired gSlider data, real-valued diffusion processing[38] was applied to each RF-encoded volume. (iii) gSlider reconstruction was then performed to obtain high slice-resolution data, using a forward model based on the Bloch simulated slab profiles of the gSlider encodings, which can be expressed as:

$$\mathbf{X} = (\mathbf{A}^T\mathbf{A} + \lambda \mathbf{I})^{-1} \mathbf{A}^T\mathbf{b}, \qquad (3)$$

where **b** is the concatenation of acquired thin-slab data at a given in-plane spatial location, **X** is the corresponding super-resolution reconstruction, **A** is the RF-encoding matrix that contains the sub-slab profiles simulated from the Bloch equations and $\lambda$ is a Tikhonov regularization parameter. To reduce slab-boundary artifacts of super-resolution images, $B_1^+$ and $T_1$ corrections[39] were incorporated into the gSlider reconstructions. After these steps, the reconstructed high slice-resolution images were used for further image analysis.

To this end, whole-brain 1 mm isotropic resolution diffusion imaging data were acquired with BUDA-gSlider-EPI. The protocol used: FOV=220×220×130 mm³, $R_{inplane}$×gSlider =4×5, 26 thin-slabs (5 mm slab-encoding), b = 1000 s/mm² with 64 diffusion-directions and 4 interleaved b =0 s/mm², TR/TE =3500/86 ms. The total acquisition time is ~40 minutes. A matching $T_2$ weighted 3D-FSE data was acquired to serve as a distortion-free reference.

The virtual-coil BUDA and gSlider reconstruction algorithms were implemented in MATLAB R2014a (The MathWorks, Inc., Natick, MA). The reconstructed data were then corrected for motion and eddy-current distortion using the "eddy" function from the FMRIB Software Library[22] (FSL, https://fsl.fmrib.ox.ac.uk/fsl/fslwiki/). Diffusion tensor model was fitted using FSL's "dtifit" function to obtain fractional anisotropy (FA) maps and primary eigenvectors.

## 2.4 Fast $T_2$ mapping using BUDA-gSlider with echo-time shuffling

High isotropic-resolution $T_2$ mapping has great potential in clinical and neuroscience applications[40] but its acquisition is hindered by low SNR and long acquisition time. In spite of recent technical advances[41–45], rapid and robust acquisition remains a challenge for whole-brain high-resolution quantitative imaging. Based on the BUDA-gSlider acquisition, we propose a model-based reconstruction which combines shuffling algorithm[45] with different echo-time (TE) to achieve distortion-free high isotropic-resolution images with different $T_2$-weighted contrasts.

As shown in Figure 4(a), three out of five different TEs were acquired in each RF-encoded slab and were individually reconstructed by VC-BUDA. The sampled thin-slab images were then combined to expand along the RF-encoding-TE dimension of $b_{exp}$ and incorporated into our joint shuffling-gSlider model. By using the extended phase graph (EPG) algorithm[46], a dictionary with signal evolution curves of $T_2$'s from 1 to 1000 ms was built (temporal basis $\Phi$ shown in Figure 4b). Figure 4(c) shows our proposed reconstruction, where the joint gSlider-shuffling model reconstructs the super-resolution slice images and exploits the low-rank property by projecting the super-resolution images along the TE dimension to temporal basis simultaneously. The thin super-resolution temporal coefficient maps $c_{exp}$ can be obtained by solving:

$$\min_{c_{exp}} \| b_{exp} - A_{exp}\Phi_{exp}c_{exp} \|_2^2 + R(c_{exp}), \quad (4)$$

where $A_{exp}$ is the expanded RF-encoding matrix, $R(c_{exp})$ is the Tikhonov regularization and $\Phi_{exp}$ is the expanded temporal basis. The weak temporal coefficients (C3 to C5) were truncated using low-rank approximation. Thin-slice images with different $T_2$ contrast were recovered using $\Phi_{exp}^T c_{exp}$ and matched to the pre-calculated dictionary to obtain the final $T_2$ maps (Figure 4c).

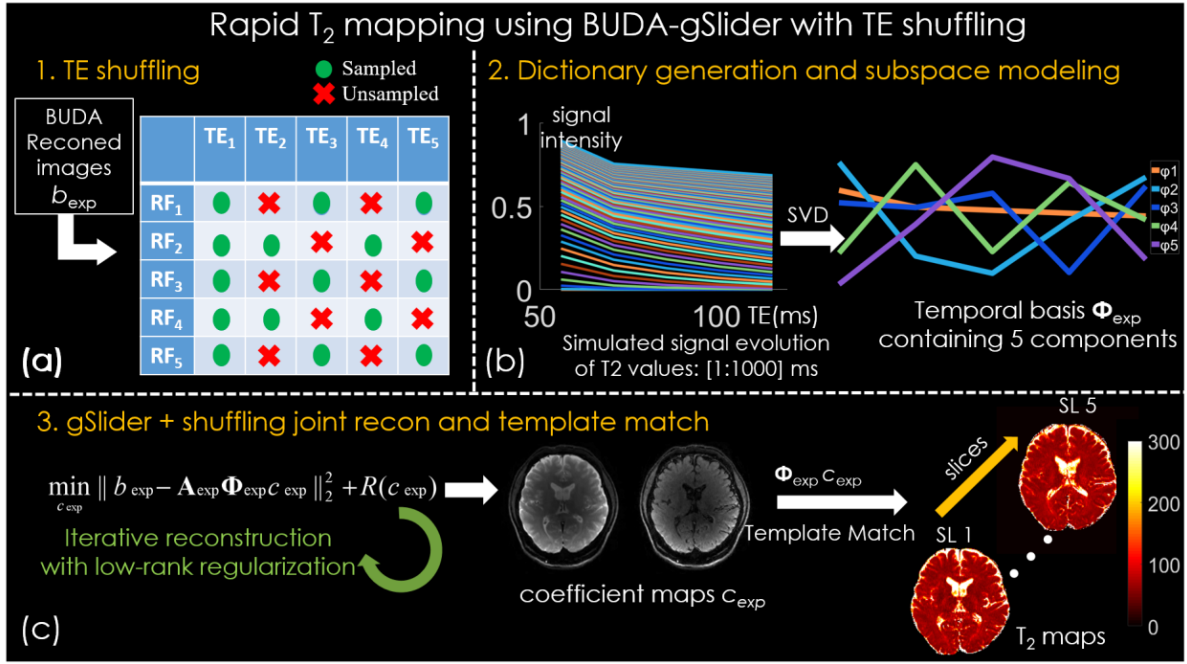

Figure 4. Reconstruction process including initialization with shuffled sampling pattern (a), dictionary generation (b), gSlider-Shuffling joint reconstruction and template match with a pre-calculated $T_2$ dictionary (c).

For rapid, high-quality $T_2$ mapping, the $T_2$ BUDA-gSlider with 2-shots at multi-band factor=2 and $R_{inplane}$=4 acceleration was acquired with 5 RF-encodings. The imaging parameters were: FOV= 220×220×130 mm$^3$, 26 thin-slabs (slab-thickness=5mm), TR=2100 ms, 5 TEs=[56, 71, 86 101 and 116] ms. With TE-shuffling acquisition, only three out of five TEs were selected per each RF encoding. The shuffled sampling pattern is shown in Figure 5(a). The proposed $T_2$-gSlider enables high-quality whole-brain $T_2$ mapping with 1-mm isotropic resolution in 63 seconds. To assess the accuracy of the estimated $T_2$ maps from BUDA-gSlider, a single-echo spin-echo sequence with seven TEs were acquired as the gold standard for comparison. The imaging protocol of single-echo spin-echo were: seven TEs=[25, 50, 75, 100, 125, 150, 200] ms, slice thickness =5 mm, in-plane resolution = 1×1 mm$^2$, TR=3500 ms, total acquisition time = 30 minutes.

### 2.5 Multi-shot EPI with wave controlled aliasing in parallel imaging

Wave controlled aliasing in parallel imaging (wave-CAIPI) employs extra sinusoidal gradient modulations during the readout to effectively use the coil sensitivity to achieve higher accelerations through parallel imaging[47,48]. Herein, we demonstrate the application of wave-CAIPI in EPI to provide higher in-plane acceleration and reduce $B_0$-related distortion[49,50]. We further combine wave-EPI with multi-shot acquisition to significantly reduce the effective echo-spacing time, and employ Hankel structured low-rank matrix completion to gain robustness against shot-to-shot phase variations.

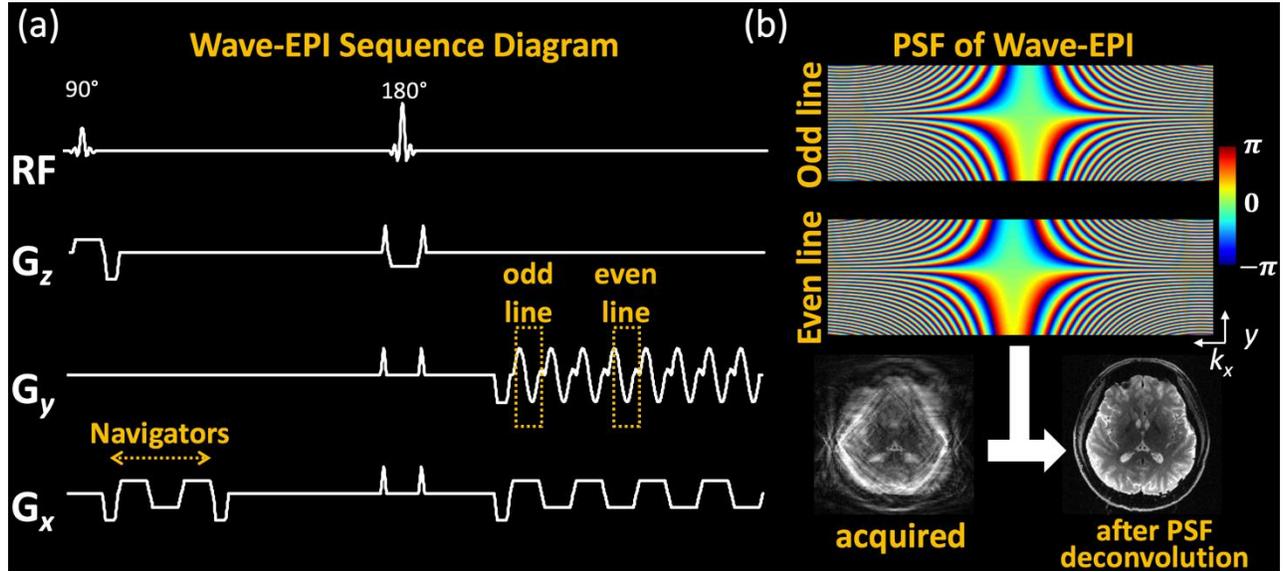

Figure 5. (a) EPI sequence diagram with wave-CAIPI. The $G_y$ gradient trajectory is time-reversed between even and odd lines to ensure that the same wave corkscrew is traced out in k-space lines with opposite polarities. (b) To account for the gradient timing differences, a separate PSF is used for the even and odd lines, where the delay along the readout ($k_x$) axis is visible. Last row demonstrates the application of PSF formalism to deconvolve the image spreading due to the sinusoidal gradients in wave-EPI and obtain a clean image.

Figure 5a shows the EPI sequence diagram with wave-CAIPI. Due to eddy currents and system imperfections, the actual k-space trajectories as captured by a point spread function (PSF) formalism differ between the even and odd lines of wave-EPI as shown in Figure 5b. To mitigate shot-to-shot variations in wave-msEPI, we enhanced the low-rank property among multiple shots by truncating the singular values of the Hankel matrix as described in **2.1**. Parallel imaging and separate even and odd PSF information are captured in a generalized SENSE forward model during the reconstruction.

To demonstrate preliminary results from wave-msEPI, we collected two acquisitions at $R_{inplane}$=12-fold acceleration using 32 channel receiver and a single cycle of sine-wave gradient: (i) 4-shot EPI at 1×1×2 mm$^3$ voxels, FOV = 220×220×80 mm$^3$ using a maximum wave gradient amplitude 15 mT/m. (ii) 3-shot acquisition at 0.8×0.8×5 mm$^3$ voxels, FOV = 220×220×150 mm$^3$. The higher in-plane resolution and lower readout bandwidth of the second acquisition allowed for higher maximum gradient amplitude of 24 mT/m.

## 3. RESULTS

Figure 6 shows the hybrid-SENSE and VC reconstruction of a slice group of 1×1×3 mm$^3$ SMS-BUDA data at multiband× $R_{inplane}$ =2×4 with partial Fourier 75% sampling. The results of individual SENSE reconstructions for blip-up and -down acquisitions show significant distortion despite $R_{inplane}$=4 acceleration. Compared to the individual SENSE, both hybrid-space SENSE and proposed VC-BUDA incorporate field maps into the forward model of the joint reconstruction to correct the geometric distortion. However, as the red arrows highlighted in Figure 6, the residual artifacts in the hybrid-SENSE reconstruction were eliminated in VC-BUDA, demonstrating that the proposed VC-BUDA has better reconstruction performance with reduced residual artifacts compared to the hybrid-SENSE.

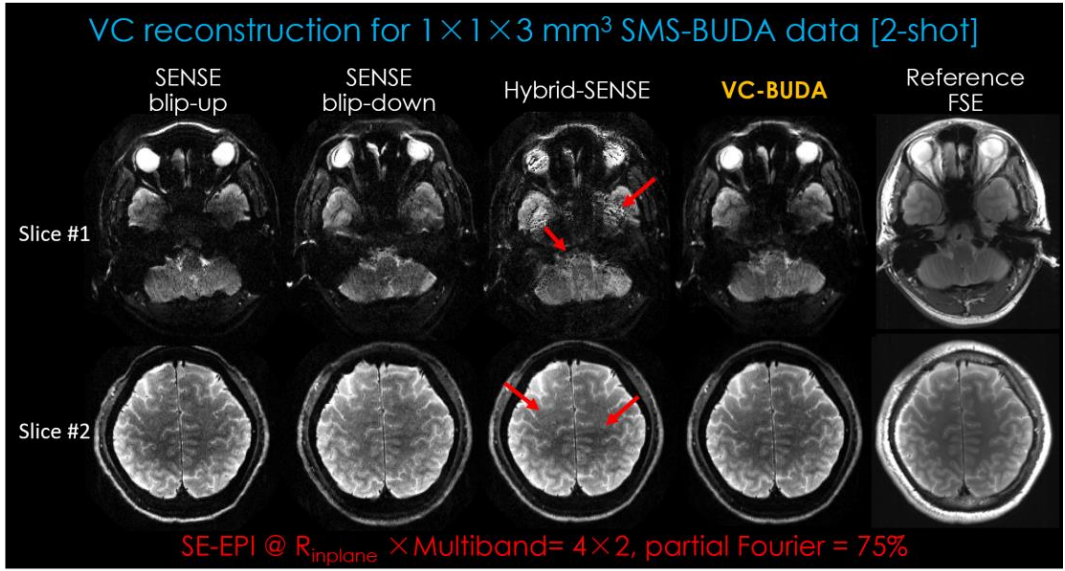

Figure 6. VC reconstruction for a slice-group of 1×1×3 mm³ SMS-BUDA data. The VC-BUDA results were compared with hybrid-SENSE and reference FSE acquisition. The red arrows indicate the residual artifacts in the hybrid-SENSE reconstruction, which were eliminated in the VC-BUDA reconstruction.

Figure 7 compares image distortion between BUDA-gSlider and reference 3D-FSE images. The whole-brain 1mm isotropic diffusion volume reconstructed by the proposed BUDA-gSlider yielded images in three orthogonal views which closely matched those of the reference $T_2$ weighted 3D-FSE.

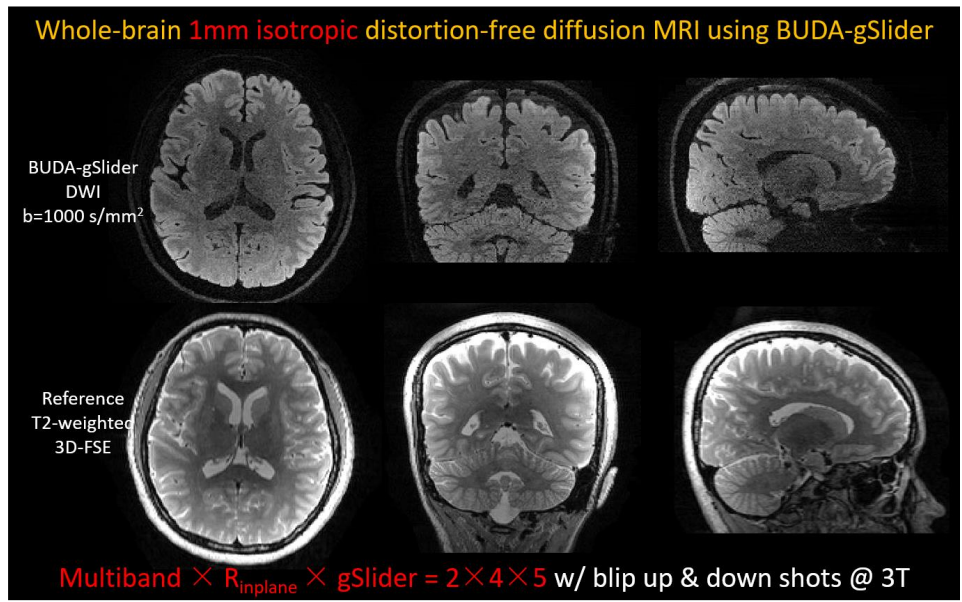

Figure 7. A 1mm-isotropic diffusion volume reconstructed by BUDA-gSlider method. The diffusion weighted volume retains very high geometric fidelity compared to the reference 3D-FSE images.

Figure 8 shows the averaged diffusion-weighted images from 64 diffusion-encoding directions and directionally-encoded color FA maps of the 1mm isotropic BUDA-gSlider diffusion data. BUDA-gSlider provides distortion-free diffusion images in the typically problematic frontal lobes, which are beneficial for mapping structural connectivity using diffusion tractography and mapping cortical diffusion patterns.

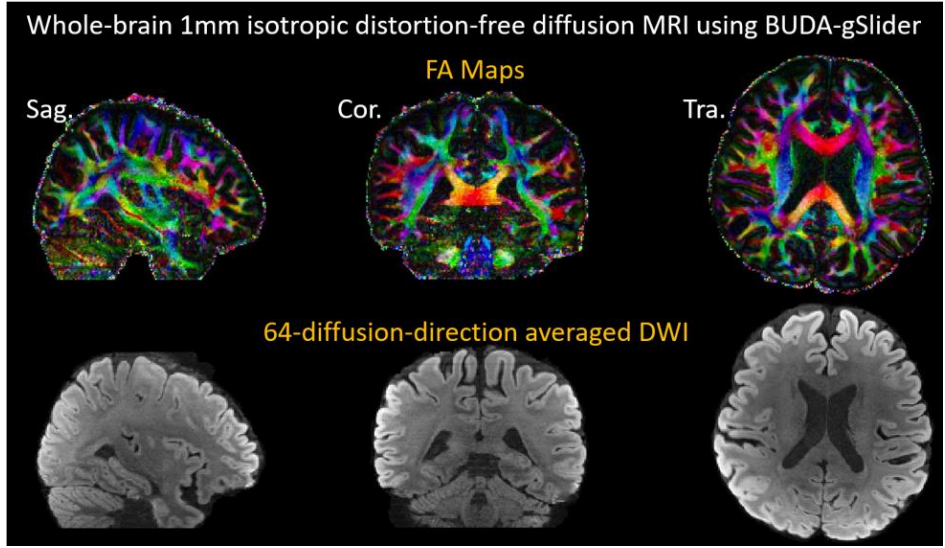

Figure 8. The FA and averaged diffusion-weighted images from 64-diffusion-direction whole-brain BUDA-gSlider data.

Figure 9 (a) shows whole-brain distortion-free $T_2$ maps from BUDA-gSlider with TE shuffling at 1mm$^3$ resolution, demonstrating high-quality whole-brain maps from a 1 minute acquisition. Figure 9 (b) compares the proposed $T_2$ BUDA-gSlider with the gold standard method. The bar plot shows the T2 values estimated by $T_2$ BUDA-gSlider are very close to the gold standard in four represent regions but ~30x faster (60s vs 30 minutes), which demonstrate the utility of the proposed fast $T_2$ BUDA-gSlider.

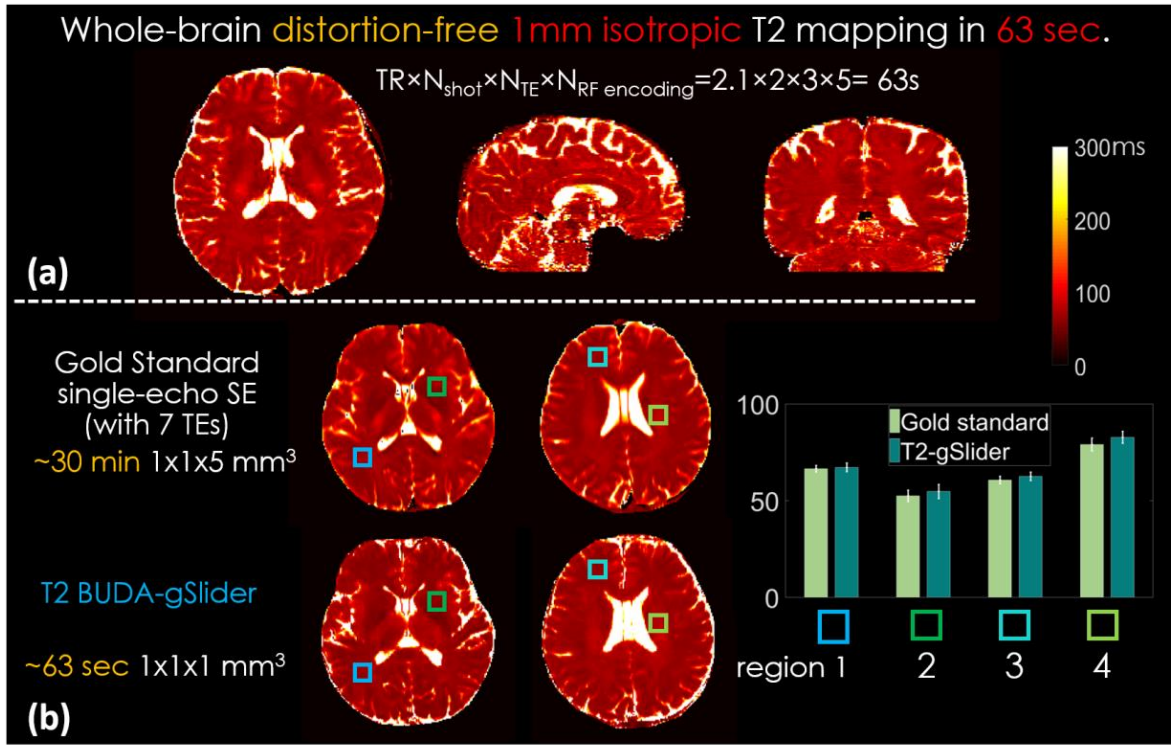

Figure 9. (a) $T_2$ maps in three views of 1-mm isotropic resolution with BUDA-gSlider acquisition with TE shuffling. (b) The comparison between $T_2$ BUDA-gSlider and gold standard method.

The preliminary results in Figure 10 demonstrate that the msEPI with wave-CAIPI could provide less aliasing artifacts (yellow arrows), less g-factor, and improved image quality compared to Cartesian msEPI. Specifically, wave-msEPI allowed for at least a 2-fold reduction in the maximum g-factor compared to the standard msEPI, and up to 30% improvement in the average g-factor metric.

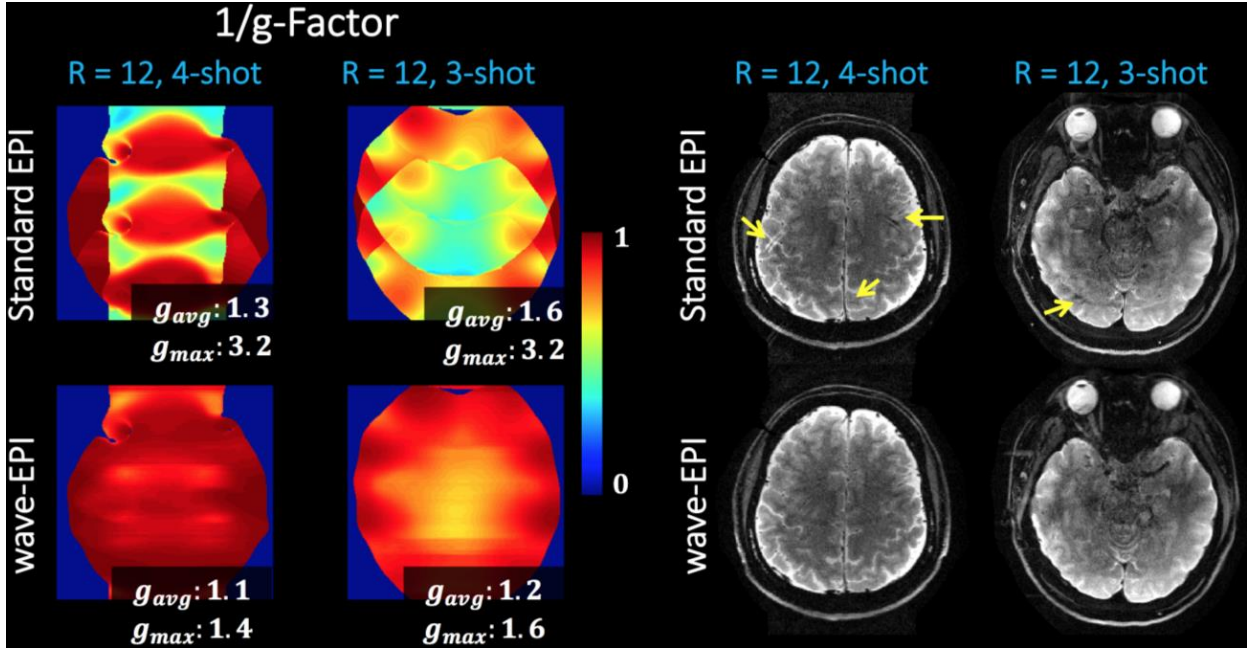

Figure 10. g-factor estimates and preliminary results for 3- and 4-shot msEPI acquisitions using standard Cartesian and wave-EPI. The yellow arrows highlight the aliasing artifacts of standard msEPI, while wave-EPI is able to mitigate the aliasing.

## 4. DISCUSSION

In this work, a highly efficient multi-shot EPI acquisition/reconstruction framework was proposed for rapid high isotropic-resolution, distortion-free structural, diffusion, and quantitative MRI. With interleaved blip up-down acquisition, the field maps were estimated and then incorporated into the forward model of reconstruction to obtain distortion-free images. To correct the shot-to-shot phase variations, Hankel structured low-rank constraint was used for multi-shot joint reconstruction. We also incorporated joint VC reconstruction and SMS concepts into BUDA reconstruction to push the limits of PF, in-plane and slice acceleration in the msEPI acquisition, and enabled high in-plane resolution, rapid structural imaging with high geometric fidelity. We further combine these methods with the gSlider acquisition to enable high SNR-efficiency, high isotropic-resolution imaging with enhanced SNR, which is beneficial for high fidelity diffusion imaging. In addition to diffusion acquisition, with TE-shuffling sampling strategy, BUDA-gSlider enables rapid whole-brain $T_2$ mapping with high isotropic resolution using SNR-efficient RF-encoded SE-EPI acquisition. Combined BUDA-gSlider reconstruction for in-plane undersampling with shuffling model for acceleration on RF-TE dimension, $T_2$ BUDA-gSlider can achieve high-quality, 1-mm isotropic whole-brain $T_2$ maps in ~1 minute without distortion. Additionally, employing wave gradients during the EPI readout could better harness sensitivity encoding to enable higher acceleration factors.

Previous studies such as hybrid-space SENSE[51] performs joint reconstruction of the 2-shots by including their phase difference and field map into the forward model to correct both shot-to-shot phase variations and geometric distortion. However, due to the low SNR of diffusion-weighted images, it is difficult to estimate the phase difference accurately, which could cause residual artifacts and noise amplification. Our proposed BUDA exploits the Hankel low-rank model-based reconstruction, and obviates the need for phase-navigation or phase difference estimation by enforcing structured low-rank prior, which provides improved image quality and SNR. Furthermore, we incorporate VC concept into the BUDA framework to further improve the reconstruction performance using the conjugate symmetry property of k-space as prior information, which is beneficial for partial Fourier reconstruction with complementary undersampling in blip-up and blip-down shots.

There are some limitations of BUDA method. Since BUDA requires an individual SENSE reconstruction for each shot to estimate the field map, inplane acceleration factor beyond 4-fold may fail to provide a robust reconstruction of individual shots. For high inplane resolution with a long EPI readout, the long $T_2^*$ decay would cause image blurring at $R_{inplane}=4$. To address these issues, we are exploring wave-EPI as well as deep learning-based methods [52,53] to achieve higher acceleration factor with robust reconstruction.

## 5. CONCLUSION

We proposed a multi-shot EPI based BUDA-gSlider acquisition/reconstruction strategy to provide high isotropic-resolution imaging without geometric distortion. We incorporated the virtual coil reconstruction and introduced wave-EPI to further improve the performance of image reconstruction. In vivo studies demonstrated that the proposed methods enabled high efficiency, high geometric fidelity multi-shot EPI for structural, diffusion weighted and quantitative imaging, which should provide high quality and efficiency MRI data in many clinical and neuroscientific applications.

## 6. ACKNOWLEDGEMENT

This work was supported in part by: National Institute of Biomedical Imaging and Bioengineering, Grant/Award Number: P41 EB015896, R01 EB017337, R01 EB019437, R01 EB020613 and U01 EB025162; National Institute of Neurological Disorders and Stroke, Grant/Award Number: K23 NS096056; National Institute of Mental Health, Grant/Award Number: R01 MH116173 and R24 MH106096; Center for Biomedical Imaging, Grant/Award Number: S10-RR023401 and S10-RR023043; NVIDIA GPU grant.